\documentclass[12pt,a4paper,english,superscriptaddress,aps,nofootinbib]{revtex4}

\usepackage{amsmath,amssymb,graphicx}

\makeatletter
\usepackage{babel}
\usepackage[active]{srcltx}
\usepackage{multirow}
\usepackage{graphicx,color}
\usepackage{changebar}
\usepackage{hyperref}
\usepackage[utf8]{inputenc}
\usepackage[T1]{fontenc}
\usepackage{ae}
\usepackage[caption = false]{subfig}

\usepackage{graphicx}
\usepackage{amsfonts}
\begin{document}

\title{Electronic states in quantum wires on a M\"{o}bius strip}


\author{J. J. L. R. Pinto}
\email[]{E-mail:joaojardel@fisica.ufc.br}
\affiliation{Universidade Federal do Cear\'{a} (UFC), Departamento do F\'{i}sica - Campus do Pici, Fortaleza-CE, 60455-760, Brazil.}

\author{J. E. G. SIlva}
\email[]{E-mail: euclides@fisica.ufc.br}
\affiliation{Universidade Federal do Cear\'{a} (UFC), Departamento do F\'{i}sica - Campus do Pici, Fortaleza-CE, 60455-760, Brazil.}

\author{C. A. S. Almeida}
\email[]{E-mail: carlos@fisica.ufc.br}
\affiliation{Universidade Federal do Cear\'{a} (UFC), Departamento do F\'{i}sica - Campus do Pici, Fortaleza-CE, 60455-760, Brazil.}

\affiliation{Institute of Cosmology, Department of Physics and Astronomy, Tufts University, Medford, Massachusetts, USA.\\}

\begin{abstract}
We study the properties of a two dimensional non-relativistic electron gas (TDEG) constrained on wires along a M\"{o}bius strip. We considered wires around the strip and along the transverse direction, across the width of the strip. For each direction, we investigate how the curvature modifies the electronic states and their corresponding energy spectrum. At the center of the strip, the wires around the surface form quantum rings whose spectrum depends on the strip radius $a$. For wires at the edge of the strip, the inner edge turns into the outer edge. Accordingly, the curvature yields localized states in the middle of the wire. Along the strip width, the effective potential exhibits a parity symmetry breaking leading to the localization of the bound state on one side of the strip.


\end{abstract}

\maketitle


\section{Introduction}

In the last decades, the investigation of the low-dimensional systems, such as the graphene, had attracted much attention due to their unconventional properties \cite{geim2007rise,Katsnelson}. Geometry  has a crucial importance in graphene systems, allowing the formation of one-dimensional quantum wires and quantum rings \cite{CastroNeto,graphenegeometry,ABring}. Moreover, the curvature of the graphene surface modifies the electronic, elastic, and thermal properties of the material \cite{gaugeingraphene,corrugated}. This new field, named \textit{curvatronics}, explores new features driven by the curvature of low-dimensional samples.
In fact, by deforming the graphene layer, strong pseudo magnetic fields due to the strain and edges states are found \cite{strain1, strain2,strain3}

Among the two-dimensional geometries proposed, a graphene M\"{o}bius strip has been studied both theoretical and experimentally \cite{guo2009mobius,wang2010theoretical,caetano2009defects, zhang2017construction,mobiusdirac,hans}.
The graphene M\"{o}bius strip is a single-sided surface built by gluing the two ends of graphene ribbons, after rotating one end by $180^{o}$ \cite{spivak}. If one starts in one edge of the M\"{o}bius strip and takes a $360^{o}$ rotation, it ends up on the other edge of the strip \cite{mobiuselasticity}. Thus, besides the curvature, the asymmetry of the M\"{o}bius strip ought to influence the electronic properties \cite{Monteiro:2023avx}. The non-trivial topology of the M\"{o}bius strip leads to remarkable properties, such as robust edge states (topological insulator) and curvature induced spin-Hall effect \cite{guo2009mobius,wang2010theoretical,caetano2009defects, zhang2017construction,mobiusdirac,hans}. 
For a graphene M\"{o}bius ring, the electron dynamics is described by the usual massless Dirac equation \cite{guo2009mobius,wang2010theoretical,caetano2009defects, zhang2017construction,mobiusdirac,hans,Monteiro:2023avx}. However, here we study the effects of the curved M\"{o}bius strip on a non-relativistic electron, describing a two dimensional electron gas (TDEG) constrained on a M\"{o}bius surface.

In the continuum limit, the influence of curvature on a constrained electron on a surfaces can be investigated by means of the so-called squeezing method, wherein 
the Schr\"{o}dinger equation on the surface is obtained by considering a small width $\epsilon$ to the surface and taking the limit $\epsilon\rightarrow 0$ \cite{dacosta}. As a result, a curvature-dependent potential is obtained, the so-called da Costa potential \cite{dacosta,jensen}. The effects of the da Costa potential have been studied on several surfaces, such as the catenoid\cite{catenoid1,catenoid2}, helicoid\cite{helix}, and nanotorus\cite{torus}, among others. Albeit simplified, the continuum quantum mechanics on surfaces is an approach capable of capture the main effects of the curved geometry on a TDEG.

The properties of nonrelativistic quantum mechanics on the M\"{o}bius strip have been previously explored in other works, for instance, in the ref. \cite{gravesen2005eigenstates,miliordos2011particle,li,Kalvoda:2019pqc} where only the minimal coupling was considered. In other words, the contribution of surface was considered only through the modification of the kinetic energy operator for curvilinear coordinates, and no geometric potential.
In this work, unlike previous studies, we consider the curvature-dependent da Costa potential in the non-relativistic electronic Hamiltonian. 
The non-relativistic Hamiltonian inherits the M\"{o}bius strip symmetries and anisotropies. As a result, the angular momentum is no longer conserved and the corresponding $\hat{L}_z$ operator does not commutes with the Hamiltonian \cite{li}. In order to overcome this issue, we follow the same approach adopted in Ref.\cite{Monteiro:2023avx}, where the electron was further constrained to wires along the M\"{o}bius strip.
The electron is restricted to move in two possible wires: one along the length of the strip or along the width of the strip. For the former, we vary the angular variable $\theta$ while keeping the width variable $u$ constant. For the latter, we fix $\theta$ and vary $u$. For each wire, we obtain the effective potential containing the curvature influence of that wire, and we analyze the wave functions and their corresponding energy levels. Additionally, the symmetries inherited or broken by the geometry on the electronic states are discussed.

Our results reveal that, for wires along the strip width, the electron can be confined on the inner or outer part of the strip, depending on the angle on the surface. Moreover, for wires along the strip length, the ground state is localized near $\theta=\pi$ for a central wire and near $\theta=0$ for wires along the strip edge. Likewise the relativistic analysis perfomed in Ref.\cite{Monteiro:2023avx}, we found that the curvature tends to form edge states along the width wires and to trap the electron around points along the length wires. Moreover, the effective curvature-dependent potentials depends on the ratio $a/L$, where $a$ is the strip radius and $L$ is its length. However, the states along the strip width are no so close to the strip edge as those found in the relativistic Hamiltonian \cite{Monteiro:2023avx}.

This paper is organized as follows: In section II, we present the Hamiltonian considered and we provide a brief review of the geometry of the M\"{o}bius strip. In section III, we define the wires around the strip, and along the width of the strip. For each wire, we studied the properties of the eigenfunctions and their respective spectrum. We also discuss some limiting cases in each wire, i.e., large width compared to the strip radius, etc. Finally, additional considerations and perspectives are presented in section IV.


\section{Electron on a M\"{o}bius Strip} 

In this section, we describe  the dynamics of a constrained electron on the surface of a M\"{o}bius strip. We employ an effective two dimensional electron gas (TDEG) approach for a nonrelativistic electron on a curved surface.

A restricted spinless electron in a surface, in the absence of external fields, is governed by the Hamiltonian \cite{dacosta,catenoid1}
\begin{align}
\label{hamiltonian1}
   \mathcal{\hat{H}}= \frac{1}{2m^*}g^{ij}\hat{P}_i\hat{P}_j  + V_{dC},
\end{align}
where $m^*$ is the effective mass of the electron, $\hat{P}_i:=- i \hbar \nabla _i$ is the momentum operator, and $V_{dc}$, the so-called da Costa potential,  
is a potential of geometric origin that corresponds to a contribution of the curvature in the Hamiltonian of the electron. The da Costa potential, depends on the Gaussian curvature $K$ and the mean curvature $H$ by \cite{dacosta,jensen}
\begin{equation}
V_ {g}=V_{dC}= -\frac{\hbar ^2}{2m^*}\left(H^2 - K\right). 
\end{equation}
The covariant derivative $\nabla_i$ of the momentum operator is given by $\nabla_iV^j := \partial_iV^j+\Gamma^j_{ik}V^k$, where $\Gamma^j_{ik}=\dfrac{g^{jm}}{2}(\partial_ig_{mk}+\partial_kg_{mi}-\partial_mg_{ik})$ are the  Christoffel symbols \cite{spivak}.
Thus, the spinless stationary Schr\"{o}ndinger equations in a curved surface can be written as 
\begin{align}
      \frac{1}{2m^*}g^{ij} \{ -\hbar^2[\partial _i \partial _j \psi - \Gamma ^{k} _{ij}\partial _k \psi]\}  
      + V_{g}\psi = E \psi.  
\end{align}

\begin{figure}[!htb]
\centering{\includegraphics[width=90mm]{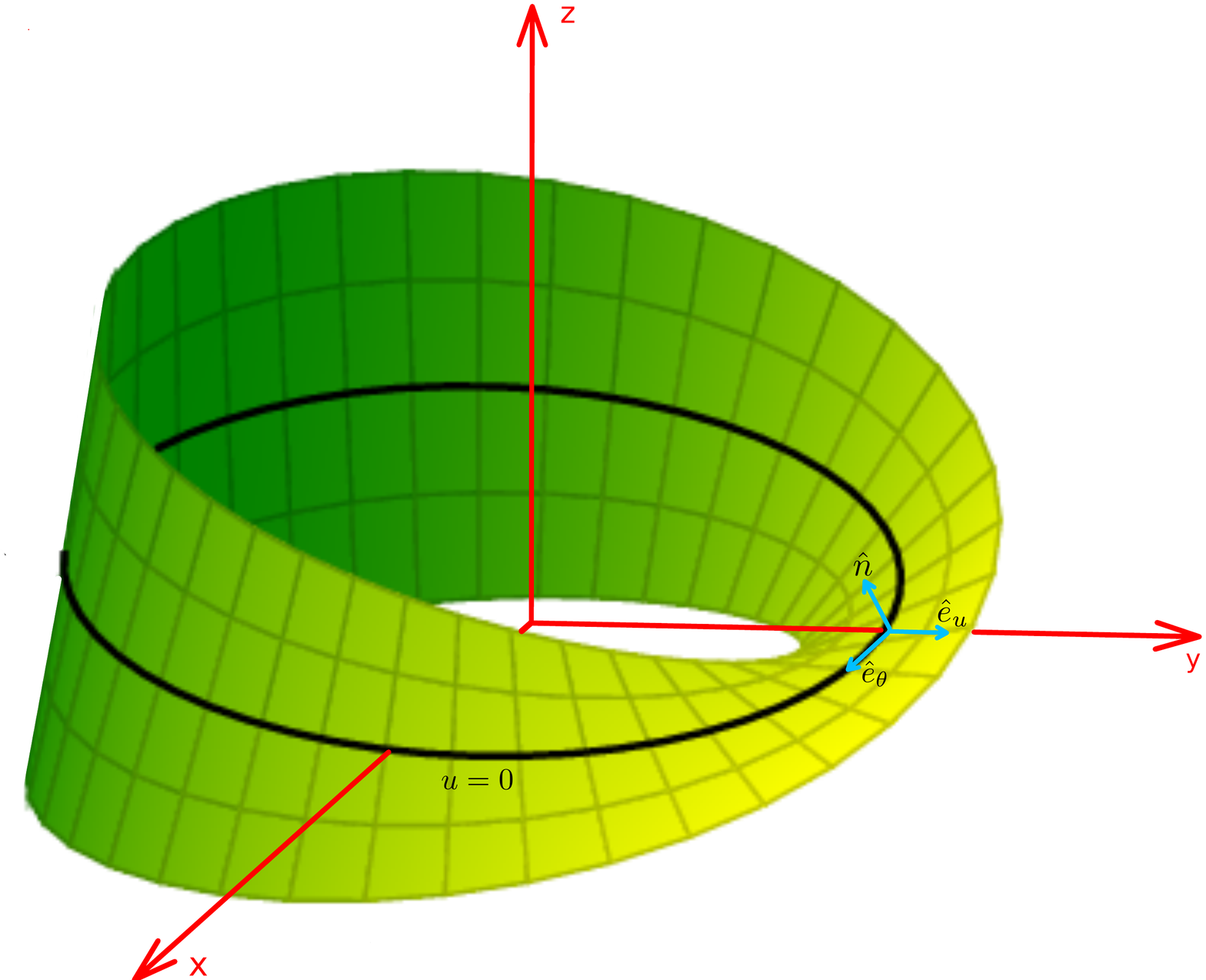}}
\caption{M\"{o}bius strip with a local reference frame determined by the tangent vectors $\hat{\mathbf{e}}_u$, $\hat{\mathbf{e}}_{\theta}$, and the normal vector $\hat{\mathbf{n}}$.
}
\label{coordinatesystem}
\end{figure}

Now let us consider the graphene M\"{o}bius strip geometry.
In cylindrical coordinates, a M\"{o}bius Strip with inner radius $a$, and width $2w$, can be described by \cite{spivak}
\begin{align}
  \Vec{\mathbf{r}}(u,\theta)= \left(a + u \cos\frac{\theta}{2}\right)\hat{\rho} +u \sin \frac{\theta}{2}\hat{k},
\end{align}
where $-w<u<w$ is a coordinate along de width strip and $0 \leq \theta \leq 2\pi$. The shape of the graphene M\"{o}bius strip is shown in fig.(\ref{coordinatesystem}).

\begin{figure}[!htb]
    \centering
    \includegraphics[scale=0.45]{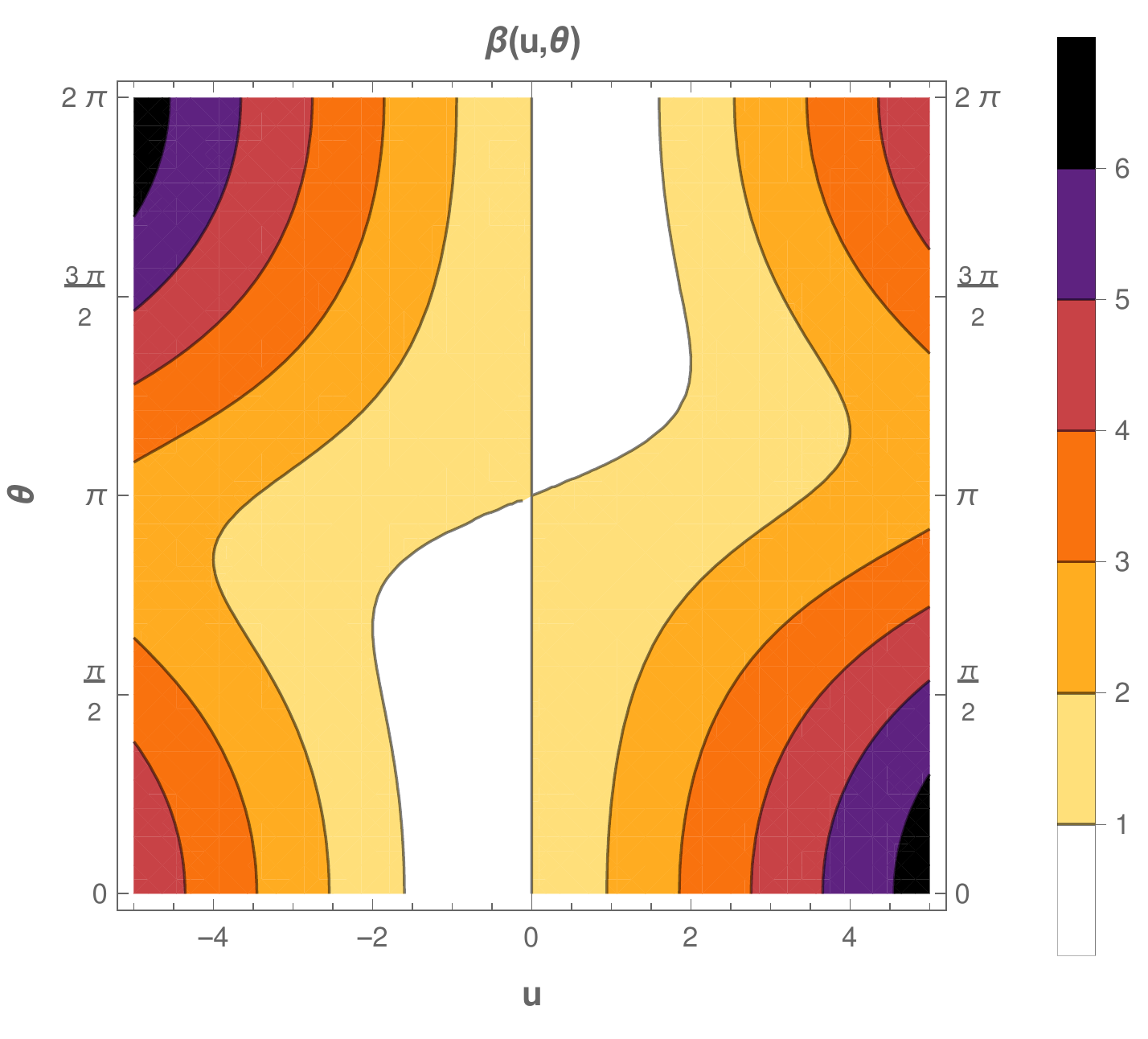}
    \caption{Metric function $\beta (u,\theta)$.}
    \label{metric}
\end{figure}

Thus, the M\"{o}bius strip metric has the form
\begin{align}
     g_{ij}=
\begin{pmatrix}
1 & 0 \\ 
0 &  \beta^2(u,\theta)
\end{pmatrix},
 \end{align}
where the angular metric component $\beta(u,\theta)$ is given by
\begin{align}
    \beta(u,\theta) = \sqrt{\frac{u^2}{4} + \Big( a + u\cos\frac{\theta}{2} \Big)^2}.
\end{align}
It is worthwhile to mention that the metric is not invariant under a parity transformation $(u,\theta)\rightarrow (-u,-\theta)$. However, $\beta$ is invariant under an inversion along the width and a rotation under a $2\pi$ angle, i.e., $\beta(-u,\theta+2\pi)=\beta(u,\theta)$. In addition, the metric is invariant under the transformation $(u,\theta)\rightarrow (-u,2\pi -\theta)$. These symmetries can be seen in fig.(\ref{metric}).

For the M\"{o}bius strip, the mean curvature has the expression \cite{spivak}
\begin{equation} H(u,\theta)=\dfrac{\sin(\theta/2)(4\beta^2 +u^2)}{8\beta^3},
\end{equation}
whereas the Gaussian curvature is given by
\begin{equation}
    K(u,\theta)=-\dfrac{a^2}{4\beta^4}.
\end{equation}
\begin{figure}[!htb]
\centering{\includegraphics[width=75mm]{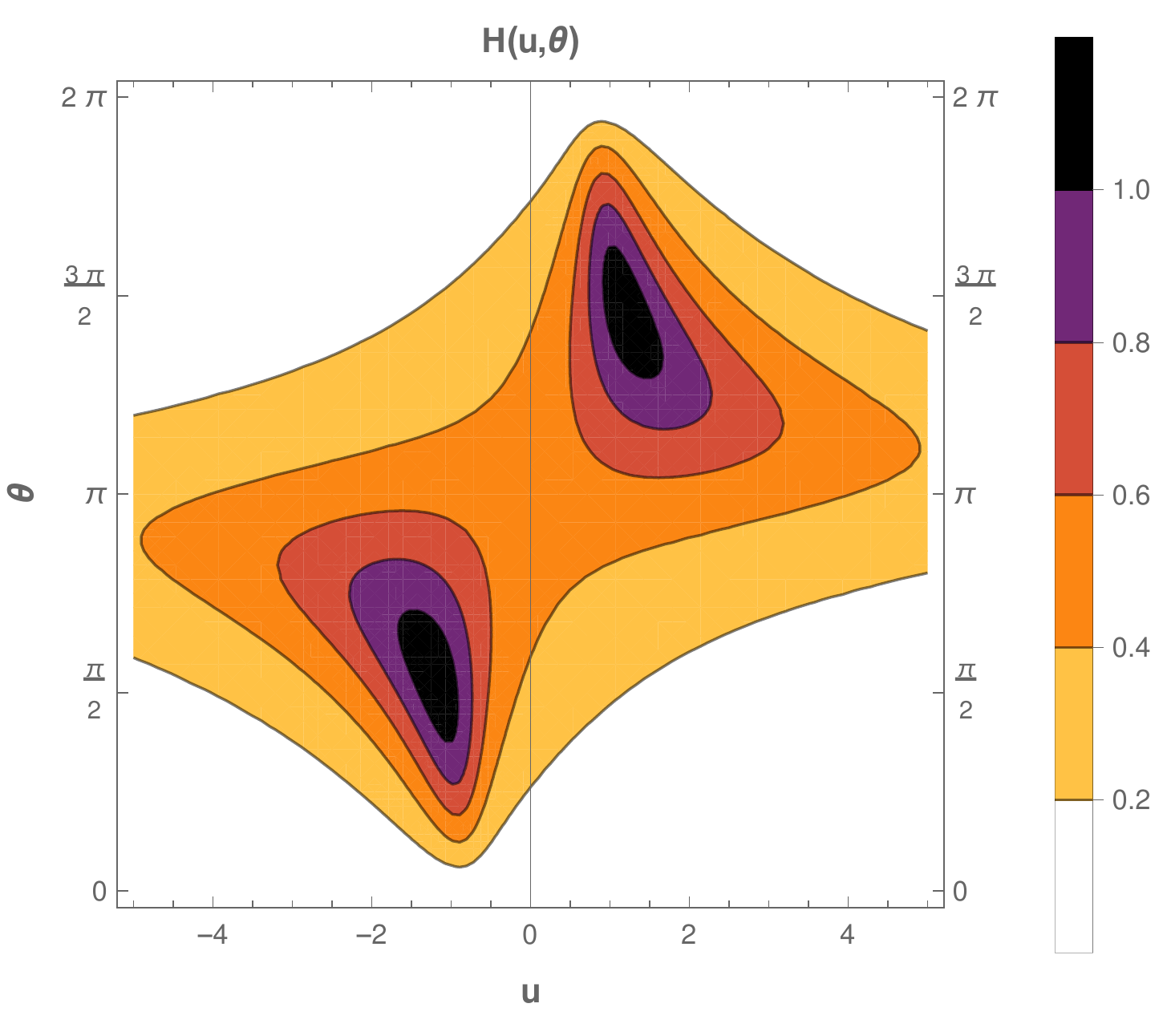}}
\caption{Mean curvature of a M\"{o}bius strip according to equation (10), where the result is varied for a  M\"{o}bius strip of internal radius $a=1$, $-5\leq u\leq 5$, and $0\leq \theta\leq2 \pi$.}
\label{meancurvature}
\end{figure}
\begin{figure}[!htb]
\centering{\includegraphics[width=75mm]{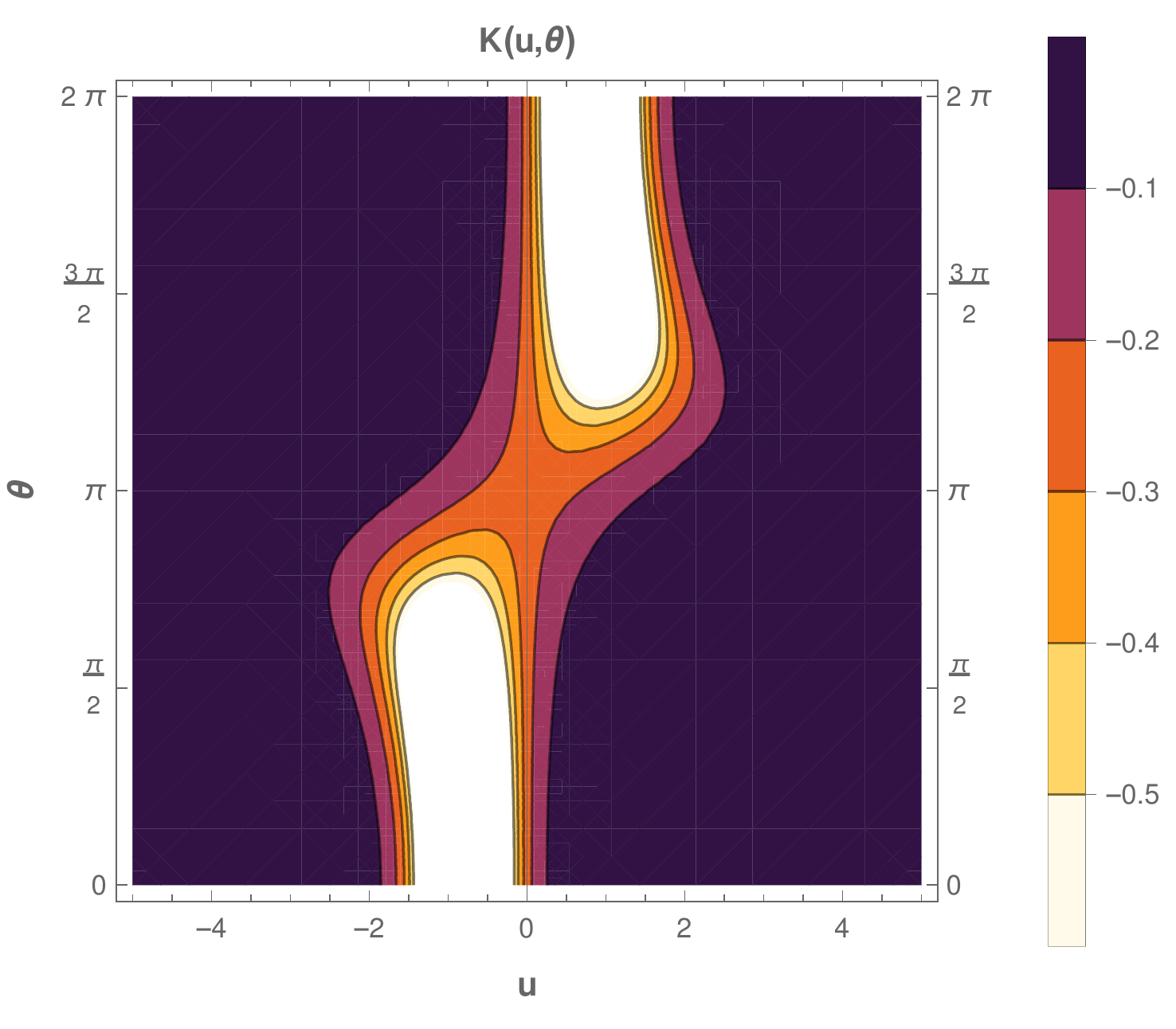}}
\caption{Gaussian curvature  of a M\"{o}bius strip according to equation (11), where the result is varied for a M\"{o}bius strip of internal radius $a=1$, $-5\leq u\leq 5$, and $0\leq \theta\leq2 \pi$.
}
\label{gaussiancurvature}
\end{figure}
The behavior of the mean curvature $H$ and the Gaussian curvature $K$
with respect to $u$ and $\theta$ is showing in the fig.(\ref{meancurvature}) and in the fig.(\ref{gaussiancurvature}), respectively. Note that the curvatures exhibit the symmetry under the transformation $(u,\theta)\rightarrow (-u,2\pi -\theta)$. Moreover, both curvatures are greater around the middle of the strip, near the angles $\theta=\pi/w$ and $\theta=\frac{3\pi}{2}$.

\begin{figure}[!htb]
    \centering
    \includegraphics[scale=0.5]{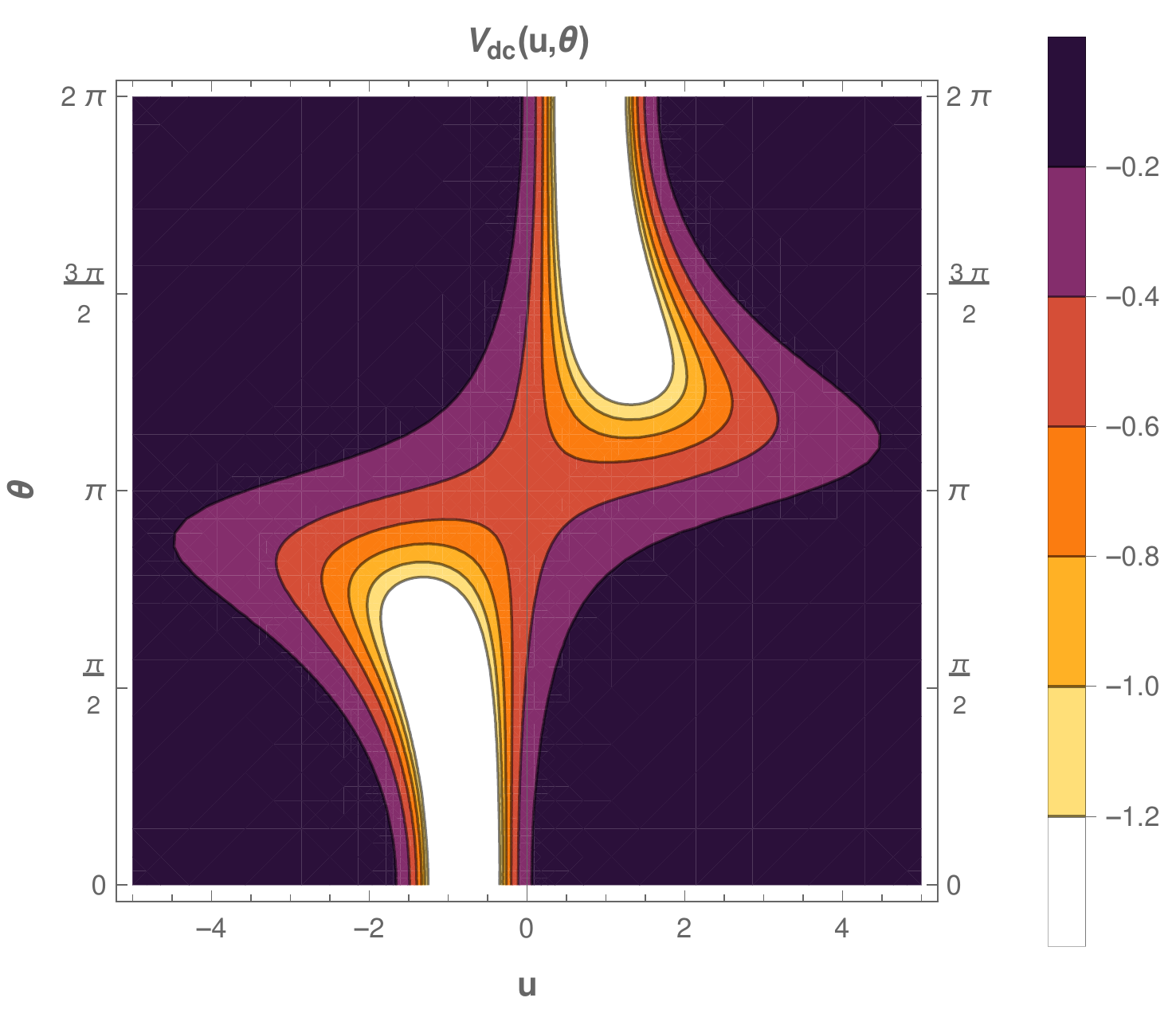}
    \caption{Geometric potential on the M\"{o}bius strip (\ref{daCosta_Potential}), where the result is varied for a  M\"{o}bius strip of internal radius $a=1$, $-5\leq u\leq 5$, and $0\leq \theta\leq2 \pi$.}
    \label{dacostapotential}
\end{figure}

Accordingly, the da Costa potential in the M\"{o}bius strip is given by
\begin{align}
\label{daCosta_Potential}
    V_{dC}=-\dfrac{\hbar^2}{2m^*}\Big[ \dfrac{\sin^2(\theta/2)}{64\beta^6}(4\beta^2 +u^2)^2 +\dfrac{a^2}{4\beta^4}\Big],
\end{align}
whose behavior is sketched in fig.(\ref{dacostapotential}).
It is worthwhile to mention that the geometric potential exhibits the same symmetries and behavior of the mean and the gaussian curvatures.

The Hamiltonian in Eq.(\ref{hamiltonian1}) can be cast into the form
\begin{equation}
\label{hamiltonian}
   \mathcal{\hat{H}}=\dfrac{1}{2\textcolor{red}{m^{*}}}\Big[\hat{P}_u^2+\dfrac{1}{\beta^2}\hat{P}^2_{\theta} -i \hbar\dfrac{\partial_u\beta}{\beta}\hat{P}_u +i\hbar\dfrac{\partial_{\theta}\beta}{\beta^3}\hat{P}_{\theta} \Big] +V_{dC},
\end{equation}
where $\hat{P}_u := -i\hbar \frac{\partial}{\partial u}$ and $\hat{P}_\theta := -i\hbar \frac{\partial}{\partial \theta}$.
It is worthwhile to mention that, unlike the catenoid, nanotubes, helicoid, and torus, the M\"{o}bius strip has no axial symmetry, i.e., the surface does not remain invariant under the transformation $\theta\rightarrow -\theta$. This asymmetry is inherited by the Hamiltonian (\ref{hamiltonian}), which now depends explicitly on $\theta$. The Hamiltonian dependence on $\theta$ means that the angular momentum with respect to the $z$ axis $L_z$ is no longer conserved. Accordingly, the $\hat{L}_z$ operator $\hat{L}_z=-\frac{i \hbar}{\beta^2}\frac{d}{d\theta}$ no longer commutes with the Hamiltonian (\ref{hamiltonian}). As a result, we cannot separate the wave function as $\Psi(u,\theta)= e^{il\theta}\psi(u)$, as was the case in the cases studied previously
\cite{catenoid1,catenoid2,helix,torus}. This issue can be overcome by freezing one of the coordinates and solving the problem for the remaining coordinate. For example, if we fix the coordinate $u$, we should obtain wires resembling rings around the M\"{o}bius Strip, whereas fixing $\theta$ would yield wires along the width of the M\"{o}bius Strip.

\section{Quantum  wires on a M\"{o}bius Strip}

In the previous section, we review the basic features of the M\"{o}bius strip geometry and explored the properties of the electron Hamiltonian. 
In this section, we define wires on the  Möbius strip where the electron can be constrained to move. We will investigate the effects of the curvature of these wires on the electronic states.

\subsection{Electron on a quantum wire around the  M\"{o}bius strip.} 

We start considering the electron constrained to move along angular wires.
By fixing the variable $u$, the stationary Schr\"{o}dinger equation $\hat{H}\Psi = E \Psi$ with Hamiltonian given by eq.(\ref{hamiltonian}) yields to
\begin{equation}
\label{angularequation}
    -\dfrac{1}{\beta^2}\dfrac{d^2\psi}{d\theta^2}+\dfrac{\partial_{\theta}\beta}{\beta^3}\dfrac{d\psi}{d\theta}-(H^2-K)\psi=\epsilon\psi,
\end{equation}
where $\epsilon=\frac{2m^{*}E}{\hbar^2}$. Although having only derivatives with respect to $\theta$, the eq.(\ref{angularequation}) depends on the choice of $u$ fixed.

By performing the change of variable
\begin{equation}
    v(u_0,\theta)=\int{\beta(u_0 , \theta)d\theta},
\end{equation}
the Schr\"{o}dinger equation becomes
\begin{align}
\label{seangular}
   -\dfrac{d^2\psi(v) }{dv^2}   
     +U(u_0,v) \psi(v) = \epsilon \psi(v). 
\end{align}
where $U(v,u_0)=-\left( H^2(u_0,v) - K(u_0,v)\right)$ is the effective potential. The variable $v$ is the arc length for a given $u$.


In addition, note that the potential is asymmetric as we change $u_0 \rightarrow -u_0$. Thus, the choice of $u_0$ leads to the localization of the electron on one side or the other side of the strip.

\begin{figure*}[!htb]
\includegraphics[width=165mm]{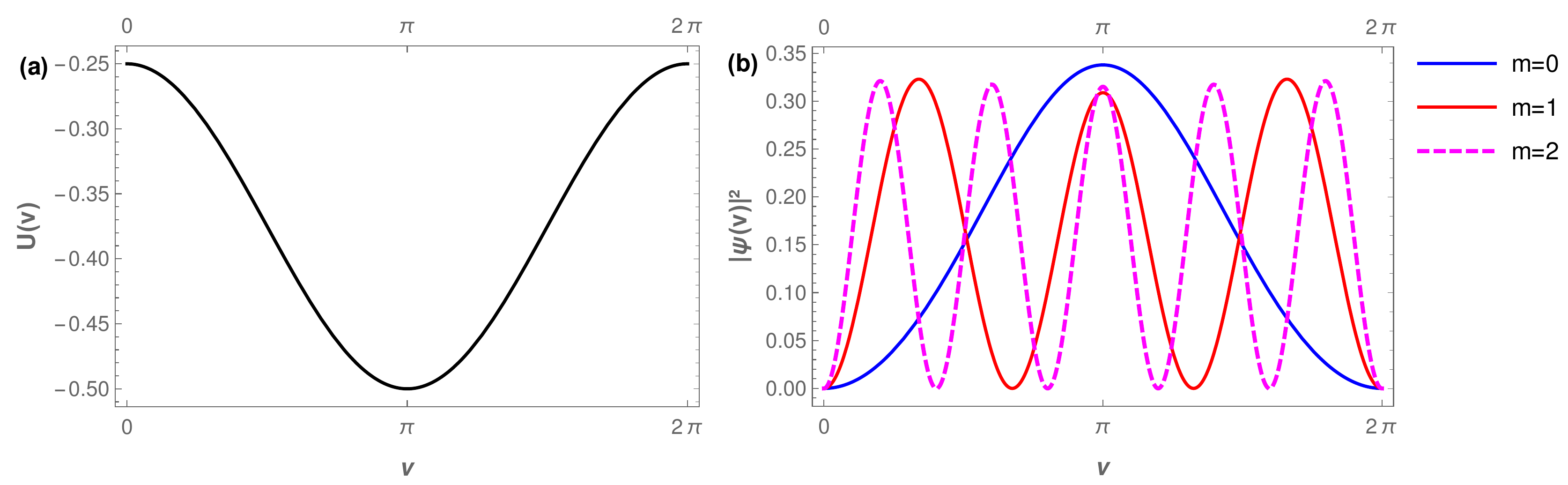}
\caption{ In graph (a) we plotted the effective potential for at the center of the M\"{o}bius strip $(u=0)$, with inner radius $a=1$ and length $L =2\pi a$. In (b) we have the probability density for the first four eigenstates of the wave function of equation (46).}
\label{angulareffectivepotential}
\end{figure*}

\subsubsection{Quantum ring at the center of the M\"{o}bius strip}

Let us start our analysis with the most symmetric case, i.e., for u=0. This wire forms a quantum ring around the M\"{o}bius strip. Indeed, for $u=0$, the metric component $\beta(0,\theta)=a$.
  
For $u=0$, the effective Schr\"{o}dinger equation eq.(18) becomes 
\begin{equation}
\label{ringequation}
  -\dfrac{d^2\psi(v)}{dv^2}+\left[-\dfrac{1}{4a^2}\sin^2\left(\dfrac{v}{2a}\right) -\dfrac{1}{4a^2}\right]\psi(v)= \epsilon\psi(v).  
\end{equation}

Thus, the effective potential for a wire on the M\"{o}bius strip for $u=0$ is given by
\begin{equation}
\label{ringpotential}
    U(v)=-\dfrac{1}{4a^2}-\dfrac{1}{4a^2}\sin^2\left(\dfrac{v}{2a}\right), 
\end{equation}
whose behavior is sketched in fig.(\ref{angulareffectivepotential}). Note that the first term in eq.(\ref{ringpotential}) represents the potential due to the curvature of the ring, whereas the second term steams from the M\"{o}bius strip curvature. 

By redefining the variable $x=\dfrac{v}{2a}$
the Schr\"{o}dinger equation along the ring at $u=0$ can be cast into a Mathieu equation
\begin{equation}
  \dfrac{d^2\psi(x)}{dx^2}+[p-2q\cos(2x)]\psi(x)= 0, 
\end{equation}
where the Mathieu parameters are given by $p= 4a^2\epsilon+\dfrac{3}{2}$ and $ q=\dfrac{1}{4}$.
Since the potential is periodic with a period of $2\pi$, we adopt the following boundary conditions
\begin{equation}
     \psi(v)=\psi(2\pi+v)
\end{equation}
Thus, the eigenstates are given by
\begin{equation}
     \psi(v)=\dfrac{1}{\sqrt{\pi}}Se_{2m+1}\left(\dfrac{v}{2a},\dfrac{1}{4}\right),
\end{equation}
where $Se_{2m+1}$ ($m=0,1,2,...$) is the Mathieu-sine function.
The energy spectrum of the particle of mass $\mu$ is related to the parameters $b_{2n+1}$ by
\begin{equation}
  E_{2n+1}=\frac{\hbar^2}{2\mu a^2}\left(\frac{b_{2n+1}}{4}-\frac{3} {8}\right),
\end{equation}
where $n=1,2,3..$ . Hence, the ground state has energy $E_0 =-\frac{3\hbar^2}{16\mu a^2}$. Compared with the usual quantum circular ring, the  wave function and energy spectrum of a particle of mass $\mu$ are:
  \begin{equation}
      \psi(\theta)=\frac{e^{im\theta}}{\sqrt{2\pi}}, \quad E^{(ring)}_m=\frac{\hbar^2}{2\mu a^2}m^2,\quad (m \in \mathbb{Z}),
  \end{equation}
where $m$ corresponds to the eigenvalue of the angular momentum of the particle in the $z$ direction \cite{li}. It is interesting to note that the spectrum of the particle in the center of the M\"{o}bius strip is not degenerate at any level, which does not occur with the particle in the circular ring, which from the ground state is doubly degenerate at each level.
We plot the probability density functions of the first three eigenstates in figure (\ref{angulareffectivepotential}).

\subsubsection{Quantum wire at the edges of the M\"{o}bius strip.}

Let us now consider the electron constrained to a wire at the edge of the strip. For a strip with a width w=2, $u$ goes from -1 to 1. The wire forms a closed loop (ring) after two complete turns around the M\"{o}bius strip, due to the $4\pi$ symmetry on the M\"{o}bius strip. The Schr\"{o}dinger equation (16), for $u=\pm 1$, is therefore
\begin{equation}
\label{schrodingerintheedges}
\begin{split}
-\dfrac{1}{\beta^2_{u=\pm 1}}\dfrac{d^2\psi}{d\theta}-\dfrac{\pm(1\pm \cos\frac{\theta}{2})\sin\frac{\theta}{2}}{2\beta^4_{u=\pm 1}}\dfrac{d\psi}{d\theta} \\
-\left[\dfrac{1}{4\beta^4_{u=\pm 1}} -\dfrac{(1+4\beta_{u=\pm 1})^2 \sin^2\frac{\theta}{2}}{64\beta^6_{u=\pm 1}}\right]\psi=\psi\epsilon,
\end{split}
\end{equation}
where $\beta^2_{u=\pm 1}=\sqrt{\frac{1}{4}+(1\pm\cos\frac{\theta}{2})^2}$.
Since the expression of the effective potential in $v$ is rather cumbersome, we present only the graphics for the potentials and their respective wave functions.

\begin{figure*}[!t]
   \includegraphics[width=160mm]{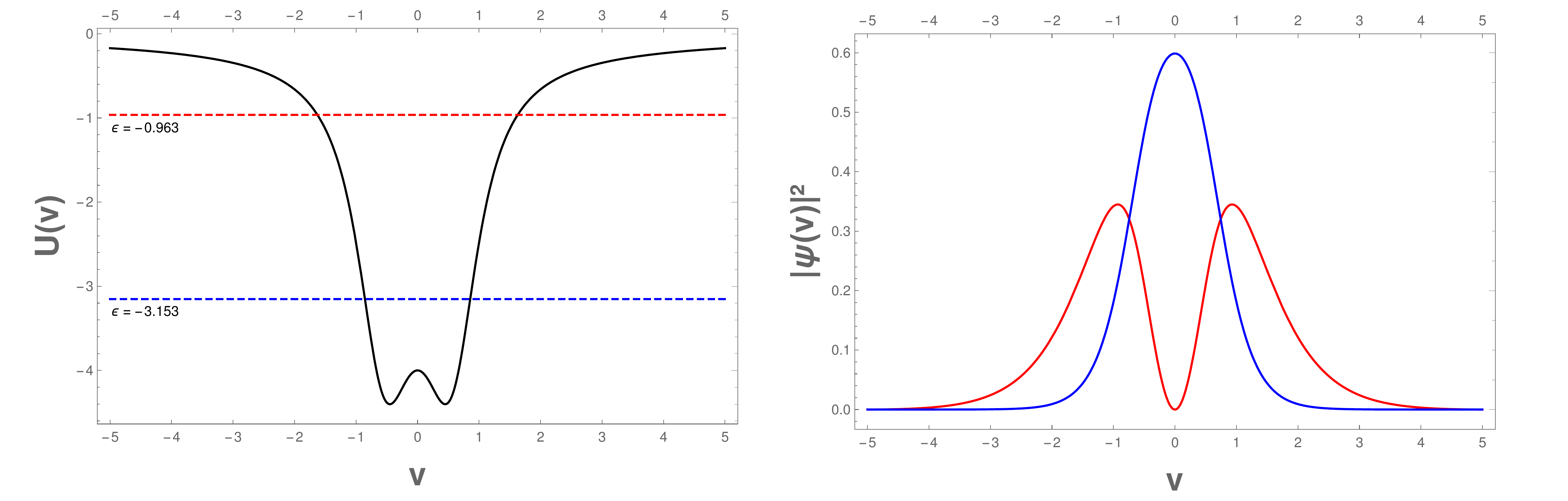}
   \caption{(a) Effective potential at the edge of a M\"{o}bius strip (solid black line). (b) The probability density function for the first bound state at the edge of the Möbius strip (solid blue line). Results for a Möbius strip with inner radius $a=1$ and length $L=5$.}
   \label{edgeground}
   \end{figure*}

In fig.(\ref{edgeground}), the effective potential is plotted for $a=1$ and $L=5$. The potential exhibits a symmetric well around the origin $v=0$. The ground state has a bell shape localized at the origin, whereas the first excited state is displaced symmetrically from the origin.

\begin{figure*}[!t]
   \includegraphics[width=160mm]{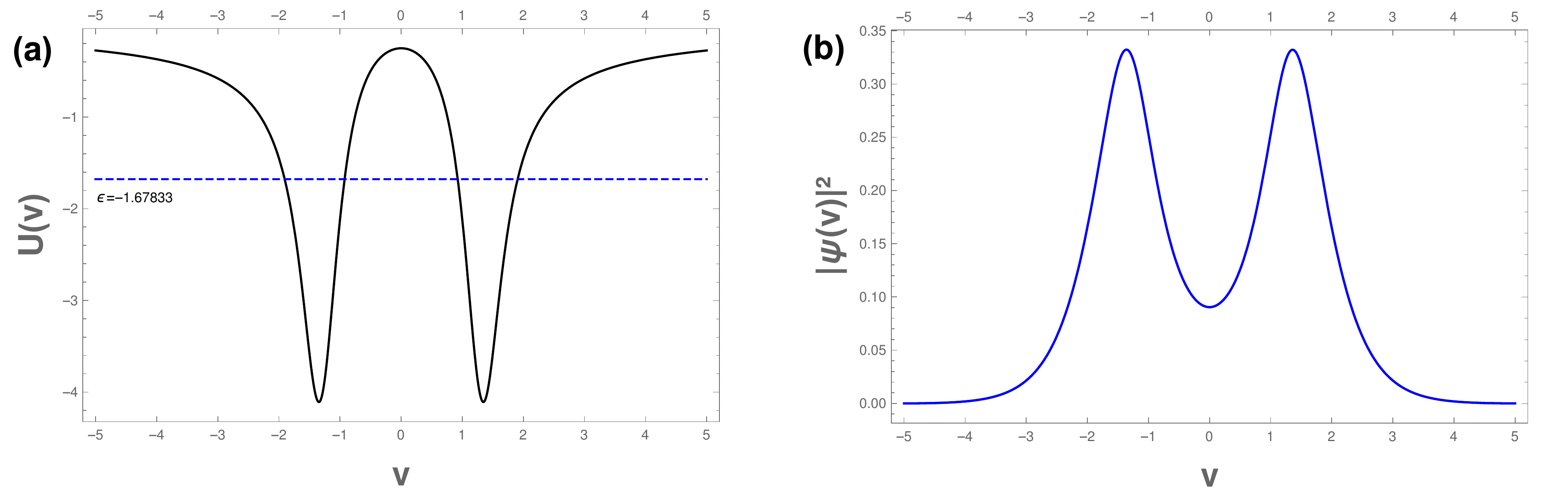} 
   \caption{(a) Effective potential at the edge of a Möbius strip (solid black line). (b) The probability density function for the ground state at the edge of the Möbius strip (solid blue line). Results for a Möbius strip with inner radius $a=0.5$ and length $L=5$.}
 \label{edgeground2}
\end{figure*}

The potential changes drastically as we reduce the inner radius $a$. In the figure (\ref{edgeground2}), we plotted the effective potential for $a=0.5$, where a barrier arises near the origin. As a result, the wave function for the ground state becomes shifted from the $v=0$.

\subsubsection{Quantum wire along the M\"{o}bius strip width}

Now let us consider the electron constrained to wires along the width strip, i.e., in the $u$ direction. We investigate how the wave function will be modified by the M\"{o}bius strip curvature.

By fixing the coordinate $\theta$, the Schr\"{o}dinger equation reads
\begin{equation}
    -\dfrac{d^2\psi}{du^2}-\dfrac{\partial_u\beta}{\beta}\dfrac{d\psi}{du}-(H^2-K)\psi=\epsilon\psi.
\end{equation}
This equation can be further simplified by considering the following change on the wave function
\begin{equation}
    \psi(u)=\psi(u,\theta_0 )=\frac{1}{\sqrt{\beta(u,\theta_0)}}\phi(u).
\end{equation}
The resulting Schr\"{o}dinger equation for the function $\phi(u)$ has the form
\begin{align}
     -\frac{d^2\phi(u)}{du^2} 
    +W(u,\theta_0)\phi(u)=\epsilon\phi(u),
\end{align}
where the effective potential is given by
\begin{align}
\begin{split}
\label{effectivepotentialu}
    W(u,\theta_0)=\left(\dfrac{\partial_u\beta}{2\beta}\right)_{|\theta=\theta_0}^2+\partial_u\left(\dfrac{\partial_u\beta}{2\beta}\right)_{|\theta=\theta_0}\\-\left( H^2(u,\theta_0) - K(u,\theta_0)\right).
    \end{split}
\end{align}
We choose as a boundary condition that the wave function vanishes at the edges of the strip,
\begin{align}
    \phi(u=-1)=\phi(u=+1)=0.
\end{align}

\begin{figure*}[!t]
\centering
   \includegraphics[width=160mm]{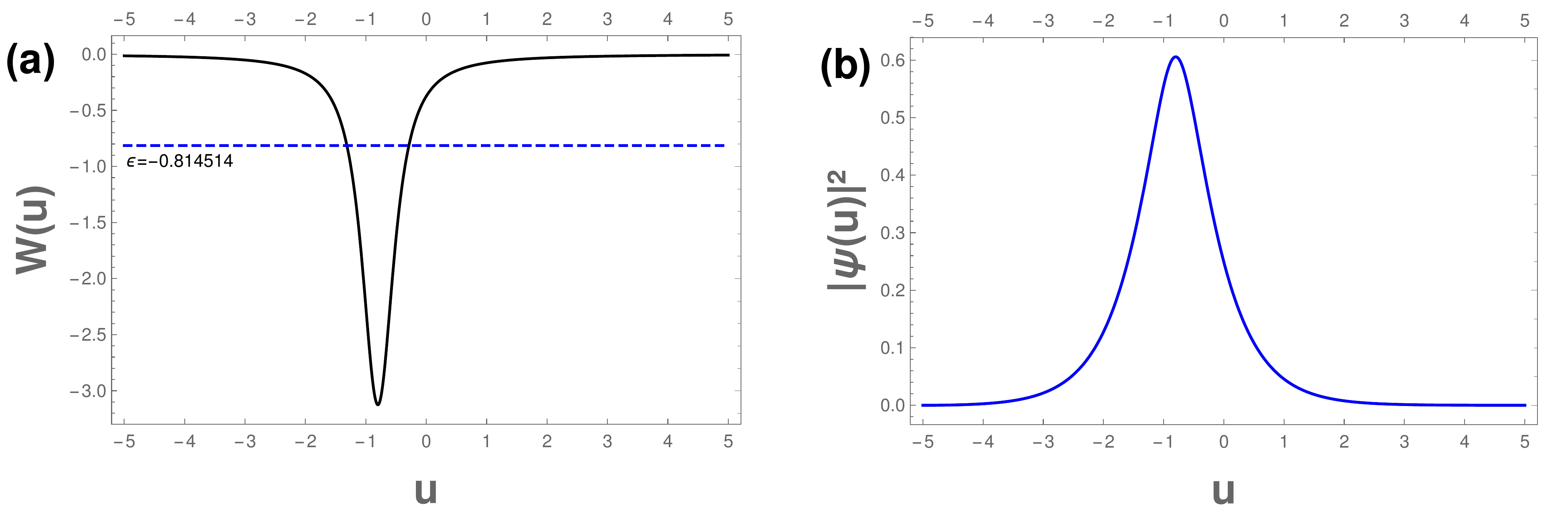}
   \caption{ (\textbf{a}) Effective potential at the width of Möbius strip for $\theta_0 =0$ (solid black line )   and energy value of the bound state
(dashed blue line). (\textbf{b})  Probability density of the bound state at the width of Möbius strip. We use $a=1$ and $w=10$.}
\label{upotential0}
\end{figure*}

\begin{figure*}[!t]
\centering
   \includegraphics[width=160mm]{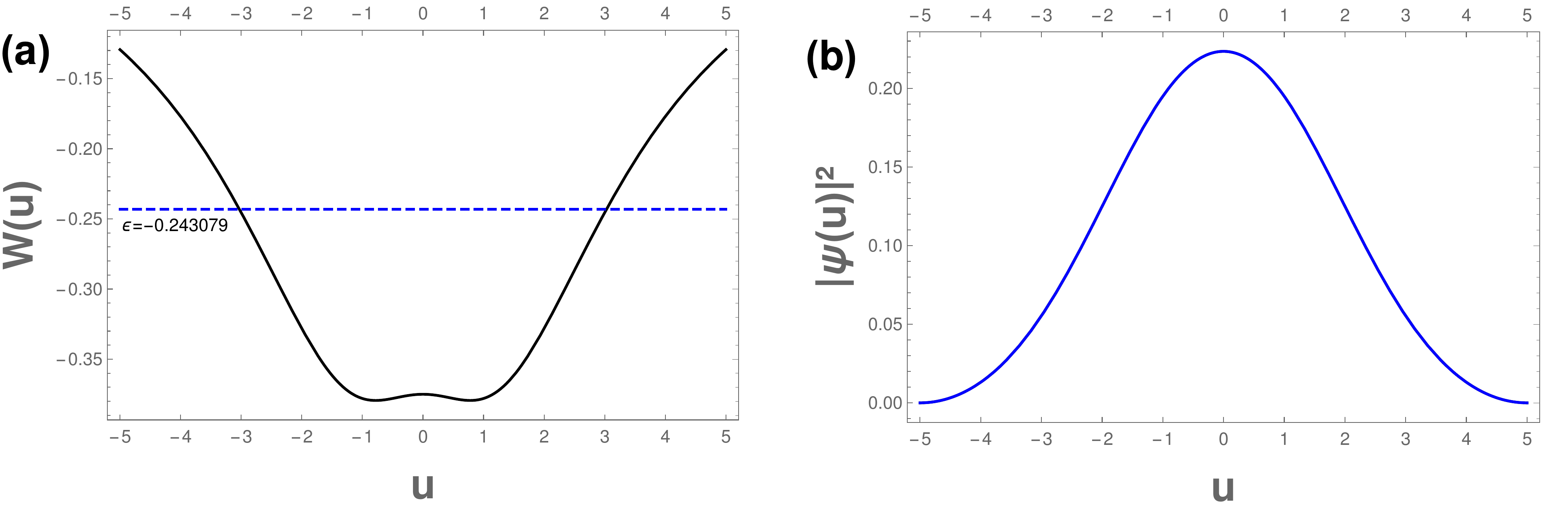}
   \caption{(\textbf{a}) Effective potential at the width of Möbius strip for $\theta_0=\pi$ (solid black line )   and energy value of the bound state
(dashed blue line). (\textbf{b})  Probability density of the bound state at the width of Möbius strip. We use $a=1$ and $w=10$.}
\label{upotentialpi}
\end{figure*}

We plot the effective potential along the $u$ direction $w(u)$ in eq.(\ref{effectivepotentialu}) for some fixed values of $\theta$. In the fig.(\ref{upotential0}), the wire is located at $\theta_0=0$. The potential exhibits a single well shifted to the left, towards the inner portion of the strip. The respective wave function is then, localized around this inner point. For $\theta_0=\pi$, shown in fig.(\ref{upotentialpi}), the potential well is symmetric with respect to the origin $u=0$. The wider potential allows the wave function to spread along the wire.

\begin{figure*}[!t]
\centering
   \includegraphics[width=160mm]{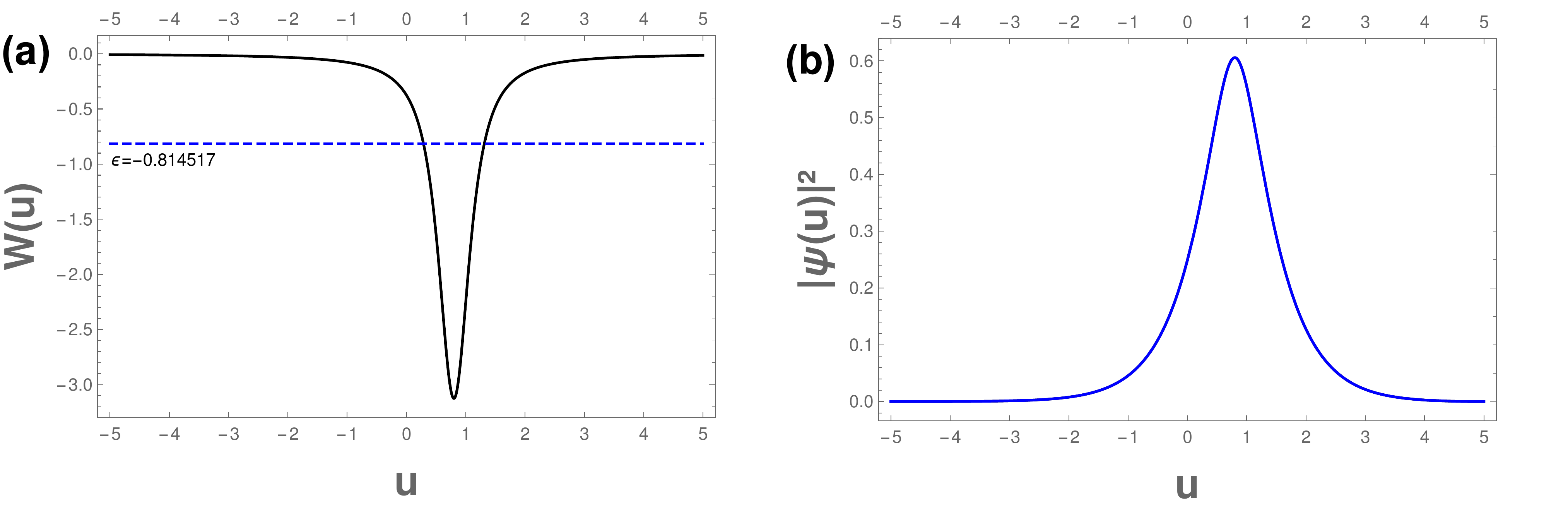}
   \caption{
(\textbf{a}) Effective potential at the width of Möbius strip for $\theta_0=2\pi$ (solid black line )   and energy value of the bound state
(dashed blue line). (\textbf{b})  Probability density of the bound state at the width of Möbius strip. We use $a=1$ and $w=10$.}
\label{upotential2pi}
\end{figure*}

Therefore, for wires along the strip width, the electron is concentrated around points which depends on the direction $\theta_0$ chosen. Note that after a $2\pi$ rotation, the wave function moves from the inner to the outer portion of the band, as expected from the symmetries of the M\"{o}bius band, which presents a parity symmetry break.

The localization of the squared wave function on the inner or outer portion of the strip, as shown in figures 9 and 11, is a feature also exhibited by the electronic states described by the Dirac equation \cite{Monteiro:2023avx}. Moreover, both in the relativistic and nonrelativistic case, the probability density has a $4\pi$ periodicity around the surface, a key property of the M\"{o}bius strip topology. However, the squared wave functions in the nonrelativistic case are not as close to the edges of the strip as those found in the relativistic dynamic \cite{Monteiro:2023avx}. The absence of true edge states might be due to the simplification in constraining the electron along the wires or a characteristic of the nonrelativistic dynamics. In any case, future analysis to clarify this point is worthy.


\section{Final remarks and perspectives}

In this work, we studied the effects of the curvature of quantum wires constrained on the M\"{o}bius strip upon the electron properties. 
The advantage of considering the wire on the surface stems from the fact that the geometric potential depends not only on the wire curvature but on the surface curvature as well. Moreover, the electronic properties are modified by the twist (topology) and the symmetries of the M\"{o}bius strip.

By considering a wire along the length of the strip at the center, i.e., for $u_0=0$, the wire forms a ring whose effective potential has a ground state localized symmetrically around $v=\pi/a$. On the other hand, the excited states are more affected by the ring curvature $1/a$. The twist of the M\"{o}bius strip also modifies the periodicity of the eigenfunctions along the length of the strip, which will have a period of $4\pi$. For an electron in a ring at the center of the M\"{o}bius strip, the spectrum and eigenfunctions are explicitly expressed in terms of the Mathieu function. Although the M\"{o}bius  geometry imposes a periodicity of $4\pi$ for the ring at its center, we obtain a wave function with a period of $2\pi$. We obtain a spectrum that depends on the ring's radius, closely resembling the case of a circular ring. However, the spectrum of the particle at the center of the M\"{o}bius strip is non-degenerate at any level, unlike the particle in the circular ring, which, starting from the ground state, is doubly degenerate at each level. These results are in agreement with the previous work in \cite{li}, although in this reference the authors did not consider the curvature-dependent da Costa potential.

 For a wire on the edge of the strip, the states are localized around $v=0$. If the inner radius $a$ is reduced compared to the strip width $w$, the ground state gets localized in two symmetrical points around the origin.

We also considered wires along the strip width, i.e., for $\theta$ fixed. The anisotropy of the M\"{o}bius strip turns the effective potential highly dependent on the chosen angle, whose dependence is transferred to the probability density. Accordingly, the electron localization strongly depends on the chosen angle. 
For $\theta_0=0$, the ground state is localized on the inner side of the strip, i.e., $u<0$. As the angle $\theta$ varies from $0$ to $2\pi$, the wave function is shifted from the inner to the outer side, i.e., for $u>0$. This is a consequence of the breaking of parity symmetry present in the geometry of the M\"{o}bius strip.
Therefore, depending on the chosen angle, electronic states can be localized either in the inner portion or the outer portion of the M\"{o}bius strip. This parity breaking property was also found in the relativistic case \cite{Monteiro:2023avx}.

Likewise the relativistic case, the nonrelativistic electronic states are strongly modified by the M\"{o}bius strip geometry, as shown in the $4\pi$ periodicity of the squared wave function. However, the tendency to form edge states (topological insulator) is not as strong as in the relativistic dynamic. Indeed, although the states are localized on the outer or inner portion of the strip for wires along the width, the electron wave functions are not so close to the strip edges as those found in the relativistic case \cite{Monteiro:2023avx}. This difference on the edge states might be due to restriction along the wires or a true property of the nonrelativistic electron on this surface. In any case, future works addressing these questions are in order.

The present work suggests further investigations. For instance, the effects of the curvature of the strip with multiple twists or the inclusion of external magnetic or electric fields could provide a way to tuning and control the density of states at other points along the strip. Moreover, the effects of the spin, by means of the the Pauli equation seem promising.

\section*{Acknowledgements}

\hspace{0.5cm}The authors thank the Conselho Nacional de Desenvolvimento Cient\'{\i}fico e Tecnol\'{o}gico (CNPq), grants n$\textsuperscript{\underline{\scriptsize o}}$ 162277/2021-0 (J.J.L.R.),  n$\textsuperscript{\underline{\scriptsize o}}$ 312356/2017-0 (JEGS),  n$\textsuperscript{\underline{\scriptsize o}}$ 309553/2021-0 (CASA) for financial support. C. A. S. Almeida is grateful to the Print-UFC CAPES program, project number 88887.837980/2023-00. C. A. S. Almeida acknowledges the Department of Physics and Astronomy at Tufts University for its warm hospitality.

\section*{Data Availability}
The datasets generated during and/or analysed during the current study are available from the corresponding author on reasonable request.

\hspace{0.5cm}

\end{document}